\documentclass[preprint,proceedings]{rmaa}

\SetYear{2002}
\SetConfTitle{Winds Bubbles and Explosions}
\title{Detection of a Collimated Jet towards a High-mass Protostar} 

\author{
  K. J. Brooks,\altaffilmark{1,2} 
  G. Garay,\altaffilmark{2}
  D. Mardones,\altaffilmark{2}
  R. P. Norris,\altaffilmark{3}}

\altaffiltext{1}{European Southern Observatory, Chile}
\altaffiltext{2}{Departamento de Astronom\'{\i}a, Universidad de Chile, Santiago, Chile}
\altaffiltext{3}{Australia Telescope National Facility, Australia}

\shortauthor{Brooks, Garay, Mardones \& Norris}
\shorttitle{Collimated Jet towards a High-mass Protostar}

\fulladdresses{
\item Kate J. Brooks: European Southern Observatory, Casilla 19001,
  Santiago 19, Chile (kbrooks@eso.org).
\item Guido Garay: Departamento de Astronom\'{\i}a, Universidad de Chile,
Casilla 36-D, Santiago, Chile.
\item Diego Mardones: Departamento de Astronom\'{\i}a, Universidad de Chile,
Casilla 36-D, Santiago, Chile. 
mardones@das.uchile.cl).
\item Ray P. Norris: Australia Telescope National Facility, P.O. Box 76,
Epping 1710 NSW, Australia. }

\listofauthors{K. J. Brooks, G. Garay, D. Mardones \& R. P. Norris}
\indexauthor{K. J. Brooks}
\indexauthor{G. Garay}
\indexauthor{D. Mardones}
\indexauthor{R. P. Norris}

\abstract{Here we present the discovery of a triple radio continuum source
 associated with IRAS 16547-4247. The spectral indices of the three
components are consistent with a jet powered by a massive O-type star in
the process of formation, with the outer radio components being the shocked
gas at the working surfaces of the jet. The detected radio continuum
emission from the central object is thought to arise from the jet itself,
prior to the formation of a detectable HII region. All three radio
continuum components are located within a molecular core of mass $10^3$
M$_{\odot}$. Our discovery makes IRAS 16547-4247 the most luminous
($\sim6.2\times10^4~L_\odot$) young stellar object from which a thermal jet
emanates, suggesting that the mechanism that produces jets in low-mass star
formation also operates in high-mass star formation. }

\resumen{}

\addkeyword{ISM: individual (IRAS 16547$-$4247)}
\addkeyword{ISM: jets and outflows}
\addkeyword{stars: formation}

\begin{document}
\maketitle

\section{Introduction}

The earliest phases of massive star birth are not well defined.  The
emerging consensus is that massive star formation begins in cold ($< 30$ K)
dense ($>10^5$~cm$^{-3}$) cores of giant molecular clouds (GMCs). Once the
stars become hot enough they ionize the surrounding gas, forming
ultra-compact HII (UCHII) regions, which then start to expand. It is not
clear if the stars themselves are formed by an accretion process similar to
that for low-mass stars or instead by collisions with lower-mass stars --
perhaps it is a combination of both (see review by Garay \& Lizano
1999). The high detection of bipolar molecular outflows in massive stars
(e.g. Beuther et al. 2002) indicates that flows are a ubiquitous phenomena
in the formation of stars across the mass spectrum. Frequently observed
towards young low-mass stars are highly collimated radio jets as well as
Herbig-Haro (HH) objects and it is now widely accepted that jets are
intimately linked to the formation of both molecular outflows and HH
objects (see review by Reipurth \& Bally 2001). The interplay between
collimated jets and molecular outflows in massive stars (B and O-type
stars) is less clear. In this higher mass regime observations of jets are
difficult, particularly because the evolutionary time scales of jets are
expected to be much shorter. Of all the known sources to have collimated
jets, the four with the brightest luminosity are: Cepheus A-HW2 at 725 pc
(Gomez et al. 1999); HH80-HH81 at 1.7 kpc (Marti, Rodriguez, \& Reipurth
1998); IRAS 20126+4104 at 1.7 kpc (Shepherd et al. 2000) and G192.16-3.82
at 1.8 kpc (Shepherd, Claussen \& Kurtz 2001). All of these objects have
luminosities less than $2 \times 10^4$ L$_{\sun}$. Up until now no such
jets have been detected towards any O-type stars.

\begin{figure*}
  \includegraphics[width=0.9\textwidth]{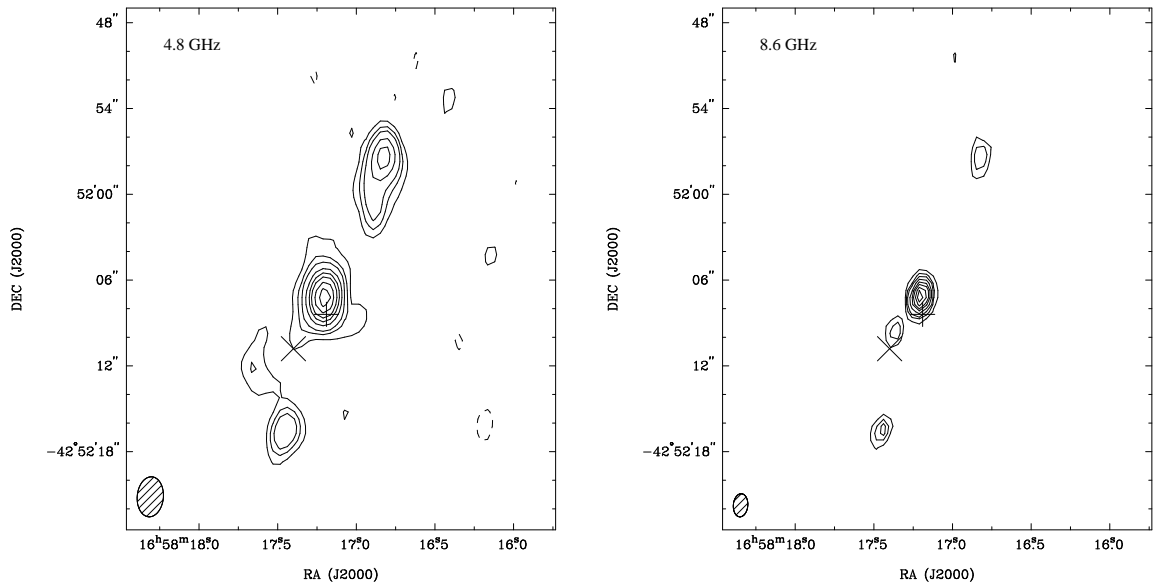} 
\caption{ATCA maps of the radio continuum emission from IRAS 16547$-$4247.
Beams are shown in the lower left corner of each panel. Contour levels are
-1, 1, 2, 3, 5, 7, 9, 12, and 15 times 0.30 mJy beam$^{-1}$ ($1\sigma$ =
0.096 mJy beam$^{-1}$ at 4.8 GHz and 0.070 mJy beam$^{-1}$ at 8.6
GHz). Also shown are the OH maser position (+) from Caswell (1998) and the
H$_2$O position ($\times$) from Forster \& Caswell (1989). }
\label{fig1}
\end{figure*}

\section{Observations}

IRAS 16547-4247 is one object from a sample of 20 luminous IRAS sources
that we are studying in detail. These sources are thought to be
representative of young massive star-forming regions. Our goal is to
understand the physical and chemical differences between different stages
of early evolution. The sources in our sample were chosen from the
Galaxy-wide survey of CS(2--1) emission towards 843 IRAS sources with
infrared colours typical of compact HII regions (Bronfman, Nyman, \& May
1996). Each source selected for our sample showed line profiles indicative
of either inward or outward motions and in some cases broad wings. Their
IRAS luminosities were in the range $2 \times 10^4 - 4
\times 10^5$ L$_{\sun}$, implying that they all contain at least one
embedded massive star ($>$ 8 M$_{\odot}$). Our study involves radio
continuum observations at four frequencies (1.4, 2.5, 4.8 and 8.6 GHz)
using the Australia Telescope Compact Array (ATCA), 1.2-mm continuum
observations using the SIMBA 37-channel bolometer array installed at
the SEST, as well as a series of SEST molecular-line observations between
85 and 250 GHz.  

\section{Results}

Fig. 1 shows the radio continuum maps at 4.8 and 8.6 GHz towards IRAS
16547-4247. The radio emission arises from three components aligned in a
southeast-northwest direction. The outer components are symmetrically
located in opposite directions from the central source, with peak positions
separated by an angular distance of $\sim$20 arcsec (0.28 pc at the
distance of 2.9 kpc, Bronfman, private communication). Using the radio data
at all four frequencies the spectral indices for each of the three
components were measured. The spectral index for the central object is 0.49
$\pm 0.12$ and is consistent with thermal emission produced by a biconical
jet (Reynolds 1986). The spectral indices of the emission from the outer
components are negative ($-0.61 \pm 0.26$ and $-0.33 \pm 0.04 $) and are
consistent with shock-induced synchrotron emission (Blandford \& Eichler 1987). This arises at the working surface of the jets, where the
collimated winds from the jet interact with the surround medium.
 
The 1.2-mm continuum emission detected towards IRAS 16547-4247 is just
resolved with a diameter of 0.4 pc and has a flux density of 16.3 Jy.
Assuming a dust opacity at 1.2 mm of 1 cm$^2$ g$^{-1}$ (Ossenkopf
\& Henning 1994) we derive a mass of $1.3 \times 10^3$ M$_{\sun}$. Results
from the line observations indicate the presence of a molecular gas core
with a molecular hydrogen column density N(H$_{2}$) of $6 \times 10^{23}$
cm$^{ -2}$, density $n$(H$_2$) of $5 \times 10^5$ cm$^{-3}$ and mass of $9
\times 10^2$ M$_{\sun}$. The triple radio source is located within the
molecular core and centered on the 1.2-mm continuum emission peak.

The total luminosity of IRAS 16547$-$4247 is estimated to be $6\times10^4
L_{\odot}$ equivalent to that of a single O8 ZAMS star.  If an O8-type star
were responsible for this high luminosity then we expect to measure a radio
flux of 3 Jy at optically thin frequencies if embedded in a uniform density
medium. However, the observed flux of the central source at 8.6 GHz is 6
mJy. One explanation for the high luminosity and weak radio emission could
be that IRAS 16547$-$4247 is a dense massive molecular core which hosts a
young massive protostar that is still undergoing an intense accretion
phase, whereby dense material is still falling towards the protostar and
quenching the development of an UCHII region (e.g. Yorke 1984).  Under this
premise the detected weak radio continuum emission from the central
component is produced by internal shocks within the jet (e.g. Anglada et
al. 1998). Adding support to this hypothesis are the characteristics of the
observed molecular line profiles. For instance, the spectral appearance of
the optically thick HCO$^{+}$(1--0) and optically thin
H$^{13}$CO$^{+}$(1--0) lines are consistent with inward-moving motions with
a characteristic infall speed of 0.75 km s$^{-1}$ implying a mass infall
rate of $\sim1\times10^{-2} M_{\odot}$ yr$^{-1}$. This is ample for the
suppression of an UCHII region around an O-type star ($\sim1\times10^{-4}
M_{\odot}$ yr$^{-1}$ according to Walmsley 1995).  Furthermore, in some
species, particularly SiO and SO, the spectra show the presence of strong
wing emission indicative of outflow activity.

IRAS 16547-4247 is associated with an MSX unresolved emission source that
is brightest at E-Band ($18.2 - 25.1$ $\mu$m). Interestingly the source
appears isolated from any bright star-forming regions. Perhaps this is why
it is possible to detect a jet here. 

\section{Conclusions}

The preponderance of the evidence suggests that we are seeing toward IRAS
16547$-$4247 a jet powered by a massive O-type star in the process of
formation, with the outer radio lobes corresponding to the shocked gas at
the working surfaces of the jet. Our discovery makes IRAS 16547-4247 the
most luminous ($\sim6.2\times10^4~L_\odot$) young stellar object from which
a thermal jet emanates, suggesting that the mechanism that produces jets in
low-mass star formation also operates in high-mass star formation.

\end{document}